\documentclass[final,5p,times,twocolumn]{elsarticle}

\usepackage{amssymb}
\usepackage{graphicx}
\usepackage{pzccal}
\usepackage{color}
\usepackage{hyperref}

\definecolor{purple}{rgb}{0.5,0,0.5}
\definecolor{blue}{rgb}{0.0,0,0.9}

\biboptions{sort&compress}

\journal{Physics Letters B}

\newcommand{\eqref}[1]{(\ref{#1})}

\begin{document}

\begin{frontmatter}

\title{Light front distribution of the chiral condensate}

\author[FZJ]{Lei Chang}
\author[ANL]{Craig D. Roberts}
\author[JARA]{Sebastian M. Schmidt}

\address[FZJ]{Institut f\"ur Kernphysik, Forschungszentrum J\"ulich, D-52425 J\"ulich, Germany}
\address[ANL]{Physics Division, Argonne National Laboratory, Argonne, Illinois 60439, USA}
\address[JARA]{Institute for Advanced Simulation, Forschungszentrum J\"ulich and JARA, D-52425 J\"ulich, Germany}

\date{9 July 2013}

\begin{abstract}
%
The pseudoscalar projection of the pion's Poincar\'e-covariant Bethe-Salpeter amplitude onto the light-front may be understood to provide the probability distribution of the chiral condensate within the pion.  Unlike the parton distribution amplitudes usually considered and as befitting a collective effect, this condensate distribution receives contributions from all Fock space components of the pion's light-front wave-function.  We compute this condensate distribution using the Dyson-Schwinger equation (DSE) framework and show the result to be a model-independent feature of quantum chromodynamics (QCD).  Our analysis establishes that this condensate is concentrated in the neighbourhood of the boundaries of the distribution's domain of support.  It thereby confirms the dominant role played by many-particle Fock states within the pion's light-front wave function in generating the chiral condensate and verifies that light-front longitudinal zero modes do not play a material role in that process.
\end{abstract}


\end{frontmatter}

\noindent\textbf{1.$\;$Introduction}.
%
%
%
%
%
The action that defines the theory of massless (chiral) quantum chromodynamics (QCD) is conformally invariant.  Associated with this feature are a dilatation current, which is conserved in a classical (unquantised) treatment of the theory, and an array of related Ward-Green-Takahashi (WGT) identities \cite{Ward:1950xp,Green:1953te,Takahashi:1957xn} between the theory's Schwinger functions.  Were these identities to remain valid in a complete treatment of the Standard Model, then the natural hadronic mass scale would be zero and all Schwinger functions would be homogeneous, with naive scaling degree.  This is plainly not the case empirically.

The conundrum is resolved by noting that classical WGT identities are derived without accounting for the effect of regularisation and renormalisation in four-dimensional quantum field theory.  This procedure leads to scale anomalies in the WGT identities originating with the dilatation current \cite{Collins:1976yq,Nielsen:1977sy,tarrach}.  Therefore, a dynamically generated mass-scale, typically denoted $\Lambda_{\rm QCD}$, is connected with \emph{quantum} chromodynamics.  The value of $\Lambda_{\rm QCD}$ must be determined empirically.

It has long been recognised that the quantum breaking of classical QCD's conformal invariance has far-reaching consequences in the analysis of high-energy processes  \cite{Brodsky:1980ny,Braun:2003rp}.  On the other hand, whilst these and related observations are instructive in principle, and motivate a class of contemporary models (see, e.g., Refs.\,\cite{Choi:2008yj,Grigoryan:2008cc,Chabysheva:2012fe,Brodsky:2013npa}), they provide little in the way of explanation for the vast array of nonperturbative strong interaction phenomena.  Knowing that scale invariance is broken by QCD dynamics is not the same as explaining how a proton, constituted from nearly massless current-quarks, itself acquires a mass $m_p \sim 1\,$GeV which is contained within a confinement domain whose radius is $r_{\rm c} \sim 1/\sigma_{\rm c}$, with $\sigma_{\rm c} \sim 0.25\,$GeV$\,\sim \Lambda_{\rm QCD}$.  Such questions can only be answered within a framework that enables the computation of bound-state properties from quantised chromodynamics.  This is highlighted further by observing that quantum electrodynamics also possesses a scale anomaly \cite{Adler:1976zt} but lies within a class of theories whose dynamical content and predictions are completely different.

Two \emph{a priori} independent, emergent mass scales are identified in the preceding passage; namely, the scale associated with QCD's confinement length and that associated with dynamical chiral symmetry breaking (DCSB), which is responsible for constituent-like behaviour of low-momentum dressed-quarks \cite{Roberts:2007ji}.  Following from the introductory discussion it appears probable that these scales are both intimately connected and originate in the same dynamics that explain the difference between the scale anomalies in QCD and QED.
However, this is not proven and the questions of whether confinement can exist without DCSB in QCD, or \emph{vice-versa}, remain open.  This fact is emphasised by the ongoing debate about coincidence of the deconfinement and chiral symmetry restoring transitions of chiral QCD in-medium 
(see, e.g., Refs.\,\cite{Braun:2011fw,Bashir:2012fs,Petreczky:2012rq,Fischer:2012vc}).

Of these two mass scales, that associated with confinement is the most problematic.  As explained elsewhere \cite{Roberts:2012sv}, there is currently no universally accepted theoretical definition of the meaning of confinement in a realistic Standard Model, which contains pions with low (lepton-like) masses.  In connection with DCSB, however, there is little ambiguity, and much can be said about its nature and role in forming hadron properties.

A fundamental expression of DCSB is the behaviour of the dressed-quark mass-function, $M(p)$ ($p$ is the dressed-quark's momentum).  It explains how an almost-massless, parton-like quark at high momentum is transformed at low momenta into a constituent-like quark that possesses an effective mass $M \sim m_p/3$.  This feature plays a critical role in forming the bulk of the visible mass in the Universe \cite{national2012Nuclear}; e.g., it ensures that the proton's mass is two orders of magnitude larger than the combined current-masses of the valence-quarks it contains.  The behaviour of $M(p)$ is also expressed in numerous aspects of the spectrum and interactions of hadrons; e.g., the large splitting between parity partners \cite{Chang:2011ei,Chen:2012qr} and the existence and location of a zero in some hadron form factors \cite{Wilson:2011aa,Cloet:2013gva}.
\smallskip

\noindent\textbf{2.$\;$Chiral Condensate}.
A derived measure of DCSB is the chiral condensate.  According to a contemporary hypothesis \cite{Brodsky:2008be,Brodsky:2009zd,Brodsky:2010xf,Glazek:2011vg,%
Chang:2011mu,Brodsky:2012ku}, this quantity is confined to the interior of the pion and hence describes intimate properties of QCD's Goldstone mode that are associated with DCSB.  From this perspective, the chiral condensate is properly defined as follows \cite{Maris:1997hd}:
\begin{equation}
\label{properqbq}
\kappa_0^\zeta = \lim_{\hat m_{u,d}\to 0}\kappa_\pi^\zeta,\quad
\kappa_\pi^\zeta := f_\pi \rho_\pi^\zeta,
\end{equation}
where $\hat m_{u,d}$ are the renormalisation-point-invariant light-quark current masses, $\hat m_{u,d} \to 0$ defines the chiral limit and
\begin{eqnarray}
i f_\pi K_\mu &=& \langle 0 | \bar d \gamma_5 \gamma_\mu u |\pi \rangle=
 {\rm tr}_{\rm CD}Z_2\!\!
\int_{dk}^\Lambda i\gamma_5\gamma_\mu
\chi_\pi(k;K)\,,\quad
\label{fpigen}\\
i\rho_\pi^\zeta &=& -\langle 0 | \bar d i\gamma_5 u |\pi \rangle=
{\rm tr}_{\rm CD}
Z_4 \!\!\int_{dk}^\Lambda \gamma_5
\chi_\pi(k;K)
\,,\label{rhogen}
\end{eqnarray}
where $K^2=-m_\pi^2$.  Here $\int_{dk}^\Lambda$ is a Poincar\'e-invariant regularisation of the four-dimensional momentum integral, with $\Lambda$ the ultraviolet regularization mass-scale, which is removed to infinity when completing any computation;
$\chi_\pi(k;K)=S_u(k_\eta) \Gamma_\pi(k;K) S_d(k_{\bar\eta})$, with $k_\eta = k + \eta K$, $k_{\bar\eta} = k - (1-\eta) K$, $\eta\in[0,1]$, and 
[with $\hat G_\pi(k;P) = k\cdot P \, G_\pi(k;P)$]
\begin{eqnarray}
\nonumber
\Gamma_\pi(k;P) & = & \gamma_5 \bigg[i E_\pi(k;P) + \gamma\cdot P F_\pi(k;P) + \\
&& +\gamma\cdot k \, \hat G_\pi(k;P) + \sigma_{\mu\nu} k_\mu P_\nu \, H_\pi(k;P)\bigg]\,, \quad\quad \label{pionBSA}
\end{eqnarray}
is the pion's 
Bethe-Salpeter amplitude;
$S_{u,d}(k)$ are the dressed light-quark propagators; and $Z_{2,4}(\zeta,\Lambda)$ are, respectively, the quark wave function and Lagrangian mass renormalisation constants, with $\zeta$ the renormalisation scale.

The content of Eq.\,(\ref{fpigen}) is well known: $f_\pi$ is the pion's leptonic decay constant, and the right-hand-side (rhs) of this equation expresses the axial-vector projection of the pion's Bethe-Salpeter wave-function onto the origin in configuration space.  Likewise, Eq.\,(\ref{rhogen}) is this wave-function's pseudoscalar projection onto the origin.  It therefore describes another type of pion decay constant.

The quantities $f_\pi$ and $\rho_\pi$ are both equivalent order parameters for DCSB; and, owing to DCSB, they are related \cite{Maris:1997hd}:
\begin{equation}
f_\pi m_\pi^2 = (m_u^\zeta + m_d^\zeta) \rho_\pi^\zeta,
\end{equation}
where $m_{u,d}^\zeta$ are the renormalisation-point-dependent current-quark masses. (N.B.\, $m^\zeta \rho_\pi^\zeta$ is renormalisation point independent and hence the ground-state pseudoscalar meson is massless in the chiral limit \cite{Holl:2004fr}.  We typically assume isospin symmetry: $m_u=m_d$.)
Furthermore, the pseudovector and pseudoscalar projections of the pion's Bethe-Salpeter wave function onto the origin in configuration space provide the only nonzero results: such projections through Dirac scalar, vector or tensor matrices yield zero.
\smallskip

\noindent\textbf{3.$\;$Light-Front Wave Functions}.
%
There is only one known approach to quantum field theory in which wave functions may be defined that possess the properties of probability amplitudes; viz., the light-front formulation, in which one obtains a Hamiltonian whose eigenfunctions are independent of the system's four-momentum \cite{Keister:1991sb,Brodsky:1997de} and hence describe states in which particle number is conserved under Lorentz boosts.  The light-front wave-function, $\varphi(x)$, of an interacting quantum system therefore provides a connection between dynamical properties of the underlying relativistic quantum field theory and notions familiar from nonrelativistic quantum mechanics.  It can translate features that arise purely through the infinitely-many-body nature of relativistic quantum field theory into images whose interpretation seems more straightforward.

Since DCSB is impossible in quantum mechanics with a finite number of degrees-of-freedom, light-front projections in quantum field theory promise a means by which to obtain quantum mechanical images of this emergent phenomenon.  That goal was achieved in Ref.\,\cite{Chang:2013pq}, which showed how to compute the pseudovector projection of the pion's Bethe-Salpeter wave function onto the light front and thereby obtain the pion's valence-quark parton distribution amplitude (PDA):
\begin{equation}
f_\pi\, \varphi_\pi(x) = {\rm tr}_{\rm CD}
Z_2 \! \int_{dk}^\Lambda \!\!
\delta(n\cdot k_\eta - x \,n\cdot K) \,\gamma_5\gamma\cdot n\, \chi_\pi(k;K) ,
\label{pionPDA}
\end{equation}
where $n$ is a light-like four-vector, $n^2=0$, $n\cdot K = -m_\pi$, and $\int_0^1\! dx \,\varphi_\pi(x)=1$.  Using Eq.\,\eqref{pionPDA}, one readily arrives at the following expression for the moments $\langle x_\varphi^m\rangle:=\int_0^1\!dx\, x^m \varphi_\pi(x)$:
\begin{equation}
f_\pi (n\cdot K)^{m+1} \langle x_\varphi^m\rangle =
{\rm tr}_{\rm CD}
Z_2 \! \int_{dk}^\Lambda \!\!
(n\cdot k_\eta)^m \,\gamma_5\gamma\cdot n\, \chi_\pi(k;K) .
\label{phimom}
\end{equation}
Plainly, one may view Eq.\,\eqref{pionPDA} as the progenitor of Eq.\,\eqref{fpigen}.

As noted above, there is another nontrivial projection of the pion wave function; i.e., the pseudoscalar projection, and on the light-front one finds \cite{Brodsky:2010xf}
\begin{equation}
\label{LFrhopi}
\rho_\pi^\zeta = \sqrt N_c \, Z_2 \int_0^1\! dx\;
\varphi_\pi(x)\, \frac{ m^\zeta }{x \bar x} + \,\mbox{instantaneous}\,,
\end{equation}
($\bar x = 1-x$) where the last term indicates contributions from the ``light-front instantaneous'' part of the quark propagator ($\sim \gamma\cdot n/k\cdot n$) and the associated gluon emission.  Given the explicit appearance of the current-quark mass, these contributions are critical to producing a nonzero chiral-limit result: one must sum infinitely many nontrivial terms in order to compensate for $\hat m \to 0$.  These nontrivial terms actually express couplings to higher Fock state components in the pion's light-front wave-function.  Such couplings are absent when one computes the $\gamma_5 \gamma\cdot n$-projection of the pion's wave function, as in Eq.\,\eqref{pionPDA} \cite{Lepage:1980fj}.  Consequently, $\rho_\pi^\zeta$ and the pseudoscalar projection of the pion's Bethe-Salpeter wave-function onto the light-front both contain essentially new information, exposing process-independent features of the pion that owe to nonvalence Fock states in its light-front wave function.  (The role of such collective behaviour in forming a chiral condensate was anticipated in Ref.\,\cite{Casher:1974xd}.)

Consider therefore the pseudoscalar projection of the pion's 
Bethe-Salpeter wave-function onto the light-front:
\begin{equation}
\rho_\pi^\zeta\, \omega_\pi(x) = {\rm tr}_{\rm CD}
Z_4 \! \int_{dk}^\Lambda \!\!
\delta(n\cdot k_\eta - x \,n\cdot K) \,\gamma_5\, \chi_\pi(k;K)\,.
\label{pionPDA5}
\end{equation}
Since the neutral pion is an eigenstate of the charge conjugation operator, $\omega_\pi(x)=\omega_\pi(\bar x)$.  We have made the notational switch $\varphi_\pi \to \omega_\pi$ in order to emphasise that the distribution amplitude defined in Eq.\,\eqref{pionPDA5} includes information about all Fock state components of the pion's light-front wave function.  In fact, with the understanding that the zeroth moment of the rhs measures the in-pion condensate, then $\omega_\pi(x)$ may be interpreted as describing the light-front distribution of the chiral condensate.   The moments of this distribution are obtained via
\begin{equation}
i \rho_\pi (n\cdot K)^{m} \langle x^m_\omega\rangle =
{\rm tr}_{\rm CD}
Z_4 \! \int_{dk}^\Lambda \!\!
(n\cdot k_\eta)^m \,\gamma_5 \chi_\pi(k;K)\,.
\label{omegamom}
\end{equation}

We would like to note that $\omega_\pi(x)$ was first considered in Ref.\,\cite{Braun:1989iv}, wherein it was identified as a twist-three two-particle distribution amplitude.  As such, it is important in the analysis of hard exclusive processes and, in particular, the study of $B$-meson pionic decays using light-cone sum rules \cite{Beneke:2003zv}.  QCD sum rules estimates of $\omega_\pi(x)$ are described in Refs.\,\cite{Ball:1998je,Ball:2006wn}.
\smallskip

\noindent\textbf{4.$\;$Asymptotic Distributions}.
Before describing a numerical computation of this distribution, it is important to develop intuition about its pointwise behaviour.  As a first step, we recapitulate upon a recent study of $\varphi_\pi(x)$ \cite{Chang:2013pq} because the same methods are applicable to $\omega_\pi(x)$.  Thus, with $\Delta_M(s) = 1/[s+M^2]$ and $\eta = 0$ in Eq.\,\eqref{pionPDA}, consider
\begin{eqnarray}
\label{pointS}
S(p) &=& [-i\gamma \cdot p + M] \Delta_M(p^2)\,, \\
\label{rhoznu}
\rho_\nu(z) &=& \frac{1}{\surd \pi}\frac{\Gamma(\nu + 3/2)}{\Gamma(\nu+1)}\,(1-z^2)^\nu\,,\\
\label{rhoEpi}
\Gamma_\pi(k;K) & = &
i\gamma_5 \frac{M^{1+2\nu}}{f_\pi} \!\! \int_{-1}^{1}\!\! \!dz \,\rho(z)
\Delta_M^\nu(k_{+z}^2)\,,
\end{eqnarray}
where $k_{\pm z} = k-(1 \mp z )K/2$.  Inserting Eqs.\,\eqref{pointS}--\eqref{rhoEpi} into Eq.\,\eqref{phimom},
using a Feynman parametrisation to combine denominators,
shifting the integration variable to isolate the integrations over Feynman parameters from that over the four-momentum $k$,
and recognising the $d^4k$-integral thus obtained as the expression for $f_\pi$, one finds
\begin{equation}
\label{momnum}
\langle x_\varphi^m \rangle_\nu =
\frac{\Gamma (2 \nu +2) \Gamma (m+\nu +1)}{\Gamma (\nu +1) \Gamma (m+2 \nu +2)}\,.
\end{equation}

Suppose that $\nu=1$.  Then $\Gamma_\pi(k^2) \sim 1/k^2$ for large relative momentum.  This is the behaviour in QCD at $k^2\gg \mu_G^2$, where $\mu_G \simeq 0.5\,$GeV is the dynamically generated gluon mass
\cite{Bowman:2004jm,Cucchieri:2011ig,Boucaud:2011ug,Ayala:2012pb,Aguilar:2012rz,Strauss:2012dg}.  $\nu=1$ in Eq.\,\eqref{momnum} yields
\begin{equation}
\label{mom1m}
\langle x_\varphi^m \rangle_1 = 
\frac{6}{(m+3)(m+2)}\,.
\end{equation}
These are the moments of
\begin{equation}
\varphi_\pi^{\rm asy}(x) = 6 x \bar x\,;
\label{PDAasymp}
\end{equation}
viz., QCD's asymptotic PDA \cite{Lepage:1980fj}.  It should be borne in mind that $\varphi_\pi(x)$ in Eq.\,\eqref{pionPDA} is actually a function of the momentum-scale $\zeta$ or, equivalently, the length-scale $\tau=1/\zeta$, which characterises the process in which the pion is involved; and that $\varphi_\pi^{\rm asy}(x)$ only provides a valid approximation to the PDA on a very small neighbourhood $\tau \Lambda_{\rm QCD} \simeq 0$ \cite{Cloet:2013tta}.

Employing an analogous procedure with Eq.\,\eqref{omegamom} yields
\begin{eqnarray}
\nonumber
\langle x^m_\omega \rangle_\nu & = & \left[m (1+m) + 2 \nu  (1+m+\nu)\right] \\
&& \times \frac{\Gamma (2+2 \nu ) \Gamma (m+\nu )}{2 \Gamma (2+\nu) \Gamma
   (2+ m+2 \nu)}\,,
\end{eqnarray}
from which one may reconstruct the distribution
\begin{eqnarray}
\nonumber
_\nu\omega_\pi(x) & = & \frac{(1+\nu) \Gamma(2+2\nu)}{2 (1+2\nu) \Gamma(\nu) \Gamma(2+\nu)} \, [x\bar x]^{\nu-1}\\
&& \times \bigg[1+
\frac{C_2^{(\nu-1/2)}(x-\bar x )}{(2\nu - 1 )(\nu+1)}  \bigg]\,, \label{omeganu}
\end{eqnarray}
where $C_2^{(\nu-1/2)}$ is a Gegenbauer polynomial of order $(\nu-1/2)$.

\begin{figure}[t]
\begin{centering}
\includegraphics[clip,width=0.9\linewidth]{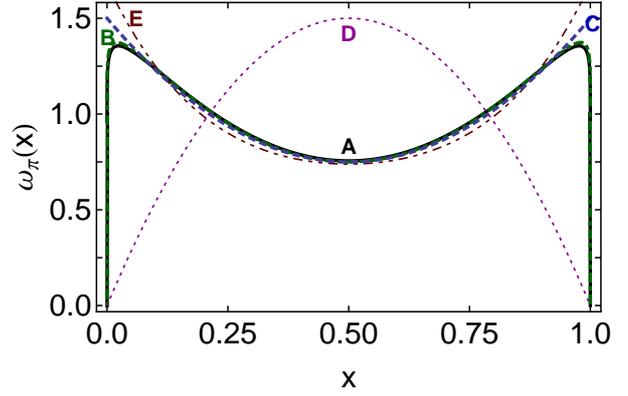}
\end{centering}
\caption{\label{Fig1}
\emph{Solid curve}\,(A) -- $\omega_\pi(x)$ computed at $\zeta_2=2\,$GeV; and \emph{dot-dashed curve}\,(B) -- $\omega_\pi(x)$ computed at $\zeta_{19}=19\,$GeV.
\emph{Dashed curve}\,(C) -- $\omega_\pi^{\rm asy}(x)$ in Eq.\,\protect\eqref{asyomega}, the asymptotic distribution of the chiral condensate within the pion.  \emph{Dotted curve}\,(D) --  for comparison, $\varphi_\pi^{\rm asy}(x)$ in Eq.\,\protect\eqref{PDAasymp}.
\emph{Dot-dot-dash-dashed curve}\,(E) -- sum rules result in Ref.\,\protect\cite{Ball:2006wn}.
}
\end{figure}

Curve-C in Fig.\ref{Fig1} is the result for $\nu=1$; i.e.,
\begin{equation}
\label{asyomega}
\omega_\pi^{\rm asy}(x) = \, _1\omega_\pi(x) = 1 + \frac{1}{2} C_2^{(1/2)}(x-\bar  x)\,.
\end{equation}
This is the asymptotic distribution of the chiral condensate within the pion, in the same sense that Eq.\,\eqref{PDAasymp} is the asymptotic form of the pion's valence-quark PDA.  The behaviour of $\omega_\pi^{\rm asy}(x)$ is striking.  It shows that the chiral condensate is primarily located in components of the pion's wave-function that express correlations with large relative momenta.

To understand this feature, recall \cite{Cloet:2013tta} that the peak in $\varphi_\pi^{\rm asy}(x)$ at $x=1/2$ is a consequence of the fact that the leading Chebyshev moment of each of the three significant scalar functions which appear in the expression for $\Gamma_\pi(k;P)$, Eq.\,\eqref{pionBSA}, occurs at $2 k_{\rm rel}:=k_\eta + k_{\bar\eta} = 0$; i.e., at zero relative momentum \cite{Maris:1997tm,Qin:2011xq} and, moreover, that these Chebyshev moments are monotonically decreasing with $k_{\rm rel}^2$.  On the other hand, owing to a quark-level Goldberger-Treiman relation; viz., for $\hat m=0$ \cite{Maris:1997hd}
\begin{equation}
\label{gtE}
f_\pi E_\pi(k;0) = B(k^2),
\end{equation}
where $B(k^2)$ is the scalar piece of the self-energy connected with the dressed quark confined within the pion, the chiral condensate may be read from the large-$k_{\rm rel}^2$ (large relative momentum) behaviour of the dominant term in the pion's Bethe-Salpeter amplitude \cite{Lane:1974he,Politzer:1976tv,Langfeld:2003ye}.
\smallskip

\noindent\textbf{5.$\;$Distribution of the Chiral Condensate}.
Having developed these insights, it is now appropriate to compute $\omega_\pi(x)$.  That can be done using the methods and the solutions of the gap and Bethe-Salpeter equations detailed in Ref.\,\cite{Chang:2013pq}.  The solutions therein were obtained with the interaction in Ref.\,\cite{Qin:2011dd}, whose infrared behaviour is consistent with that determined in modern studies of QCD's gauge sector, which indicate that the gluon propagator is a bounded, regular function of spacelike momenta, $q^2$, that achieves its maximum value on this domain at $q^2=0$ \cite{Bowman:2004jm,Cucchieri:2011ig,Boucaud:2011ug,Ayala:2012pb,Aguilar:2012rz,Strauss:2012dg}, and the dressed-quark-gluon vertex does not possess any structure which can qualitatively alter this behaviour \cite{Skullerud:2003qu,Bhagwat:2004kj}.  The interaction also preserves the one-loop renormalisation group behaviour of QCD so that, e.g., the quark mass-function is independent of the renormalisation point.

We will capitalise, too, on the fact that in completing the gap and Bethe-Salpeter kernels, Ref.\,\cite{Chang:2013pq} used two different procedures and compared the results: rainbow-ladder truncation (RL), the most widely used Dyson-Schwinger equation (DSE) computational scheme in hadron physics, detailed in App.\,A.1 of Ref.\,\cite{Chang:2012cc}; and the DCSB-improved (DB) kernels detailed in App.\,A.2 of Ref.\,\cite{Chang:2012cc}, which are the most refined kernels currently available.  Both schemes are symmetry-preserving, and hence ensure Eq.\,\eqref{gtE}, but the latter incorporates essentially nonperturbative effects associated with DCSB into the kernels, which are omitted in rainbow-ladder truncation and any stepwise improvement thereof \cite{Chang:2009zb}.

Finally, a note on the methods of Ref.\,\cite{Chang:2013pq} is appropriate.  They may fairly be described as a refinement of the spectral representation techniques explained in Ref.\,\cite{Nakanishi:1963zz}.  The numerical solutions of the gap and Bethe-Salpeter equations, obtained as matrices, are each re-expressed in terms of one or more generalised spectral functions and associated denominators that are some power of a quadratic form in the relevant spacelike momentum.  The paramount strength of this approach is that it solves the practical problem of continuing from Euclidean metric, which is the most widely used framework for practical nonperturbative studies of QCD, to Minkowski space, which is where the light-front is defined.

One detail of the generalised spectral representations is important herein.  DSE kernels that preserve the one-loop renormalisation group behaviour of QCD will necessarily generate propagators and Bethe-Salpeter amplitudes with a nonzero anomalous dimension $\gamma_F$, where $F$ labels the object concerned.  Consequently, the spectral representation must be capable of describing functions of $\mathpzc{s}=p^2/\Lambda_{\rm QCD}^2$ that exhibit $\ln^{-\gamma_F}[\mathpzc{s}]$ behaviour for $\mathpzc{s}\gg 1$.  This is readily achieved by noting that
\begin{equation}
\ln^{-\gamma_F} [D(\mathpzc{s})]
= \frac{1}{\Gamma(\gamma_F)} \int_0^\infty \! dz\, z^{\gamma_F-1}
\frac{1}{[D(\mathpzc{s})]^z}\,,
\end{equation}
where $D(\mathpzc{s})$ is some function.  Such a factor can be multiplied into any existing spectral representation in order to achieve the required ultraviolet behaviour.  In the present context, it is the anomalous dimension of the dressed-quark mass-function that must properly be represented.  Owing to Eq.\,\eqref{gtE}, this affects the pion's Bethe-Salpeter amplitude, too.

We use the procedures described above to compute the moments in Eq.\,\eqref{omegamom} and therefrom reconstruct the pseudoscalar projection of the pion's Bethe-Salpeter amplitude onto the light-front.  The $\hat m=0$ result, curve-A in Fig.\,\ref{Fig1}, is
\begin{equation}
\omega(x;\zeta_2) =
N_\nu [ x \bar x ]^{\nu-1}
[1+ a_2 C_2^{(\nu-1/2)}(x - \bar x)]\,,
\end{equation}
with $N_\nu=\Gamma[2 \nu]/\Gamma[\nu]^2$, $\nu=1.05$, $a_2=0.48$.
To reconstruct $\omega_\pi(x)$, we typically used $50$ moments.  It is straightforward but unnecessary to use more: the same distribution is obtained from $100$ moments.  The reconstruction was achieved using the Gegenbauer-polynomial method introduced in Ref.\,\cite{Chang:2013pq} and elucidated further in Ref.\,\cite{Cloet:2013tta}.  There is no ambiguity in the result, since the polynomials $\{C_j^{(\nu-1/2)}(x-\bar x),j=1,\ldots,\infty\}$  are a complete orthonormal set on $x\in[0,1]$ with respect to the measure $[x \bar x]^{\nu-1}$.  Owing to the fact that $\omega_\pi(x)=\omega_\pi(\bar x)$, only polynomials of even degree contribute.  The reconstruction procedure converges at the first step; i.e., one can terminate the series at $j=2$.  Including the second term, $j=4$, alters nothing by more than $0.1$\%.

The result in Fig.\,\ref{Fig1} is striking and we'll now explain why.
The first thing of which to be aware is that only the $E_\pi(k;K)$ term in Eq.\,\eqref{pionBSA} provides a nonzero contribution when one removes the regularisation scale $\Lambda\to\infty$.  This is because $\lim_{\Lambda\to\infty} Z_4(\zeta,\Lambda) = 0$; and whilst the integral of the $E_\pi(k;K)$ term diverges with $\Lambda$ at precisely the rate required to produce a finite, nonzero, $\Lambda$-independent result, the terms $F_\pi(k;K)$, $G_\pi(k;K)$, $H_\pi(k;K)$ provide contributions to the integral that are finite as $\Lambda\to\infty$ and hence disappear when multiplied by the vanishing renormalisation constant.

This re-emphasises the explanation we provided for Eq.\,\eqref{asyomega}.  It also entails that the result is a model-independent feature of QCD.  Since the integral is dominated by the ultraviolet behaviour of the integrand, no difference in Bethe-Salpeter kernels at infrared momenta can have an impact.  Owing to Eq.\,\eqref{gtE}, the chiral-limit result is completely determined by the momentum-dependence of the scalar piece of the self-energy associated with the dressed-quark that is confined within the pion.  This momentum-dependence is the same in all computational schemes that preserve the one-loop renormalisation group properties of QCD.  We confirmed that by computing $\omega_\pi(x)$ in both RL and DB truncation and verifying that the results are identical.

We also computed $\omega_\pi(x)$ at two different scales; viz., $\zeta_2$ and $\zeta_{19}$.  (The latter value is used commonly in DSE studies that follow Ref.\,\cite{Maris:1997tm}).  As one should expect and is evident in Fig.\,\ref{Fig1}, $\omega_\pi(x) \to \omega_\pi^{\rm asy}(x)$ as $\Lambda_{\rm QCD}/\zeta \to 0$.  However, as highlighted elsewhere \cite{Cloet:2013tta} in connection with the evolution of $\varphi_\pi(x)$, the rate of approach to the asymptotic form is extremely slow.

With the results in Fig.\,\ref{Fig1} we have provided a model-independent demonstration that the chiral condensate in Eq.\,\eqref{properqbq} is primarily located in components of the pion's wave-function that express correlations with large relative momenta, a feature which entails that light-front longitudinal zero modes do not play a material role in forming the chiral condensate.  This consequence may be elucidated by noting that $\omega_\pi(x=0)=0=\omega_\pi(x=1)$ at any finite renormalisation scale and $\omega_\pi(x) = \omega_\pi(\bar x)$.  Hence, the maximal contribution to the chiral condensate is obtained when half the partons carry a near-zero fraction of the pion's light-front momentum but the other half carry a near-unit fraction.  This discourse complements arguments to the same effect in Ref.\,\cite{Brodsky:2010xf}.
\smallskip

\noindent\textbf{6.$\;$Epilogue}.
Our prediction for the pseudoscalar projection of the pion's Bethe-Salpeter amplitude onto the light-front
is a model-independent feature of QCD.  It should be verified.  This is a theoretical challenge because few contemporary techniques with a veracious connection to QCD can provide access to anything other than the pion's valence-quark parton distribution amplitude, whereas the pseudoscalar projection
receives contributions from all Fock-states in the pion's light-front wave function.
Lattice-QCD is one applicable tool.  However, with existing algorithms it can only be used to compute one nontrivial moment of $\omega_\pi(x)$.  QCD sum rules might also be employed usefully: indeed, estimates exist \cite{Ball:1998je,Ball:2006wn}.  They, too, typically work with moments of the distribution.  Therefore, in order to assist practitioners in meeting the theoretical challenge, we present our prediction for the lowest three moments:
\begin{equation}
\label{latticemoment}
\int_0^1 \!dx\, (2 x - 1)^{2 j} \, \omega_\pi(x,\zeta_2)
= \left\{
\begin{array}{l}
 0.39\,, j=1\\
 0.25\,, j=2\\
 0.18\,, j=3
 \end{array}\right. ;
\end{equation}
and note in addition that $\omega_\pi(1/2,\zeta_2)=0.76$ cf.\ $\omega_\pi^{\rm asy}(1/2)=3/4$.

A comparison with the contemporary sum rules estimate \cite{Ball:2006wn} is worthwhile.  That result corresponds to a renormalisation scale of $\zeta_1=1\,$GeV.  It is plotted as curve-E in Fig.\,\,\ref{Fig1}, produces $j=1,2,3$ moments $0.41$, $0.27$, $0.20$, respectively, and $\omega_\pi^{\rm SR}(1/2,\zeta_1)=0.74$.  The agreement with our prediction is plainly very good.  Differences are only marked in the neighbourhood of the endpoints, something one might have anticipated given that just low-order moments can practically be constrained in a sum rules analysis and such moments possess little sensitivity to the behaviour of $\omega_\pi$ in the neighbourhood of the endpoints.  We judge that the generally good agreement with our prediction from such limited input provides strong support for the model-independent nature of our result.  This is further emphasised by the fact that the estimate in Ref.\,\cite{Ball:2006wn} improves over an earlier calculation \cite{Ball:1998je} and, as gauged by the $L^1$-norm, the modern refinement shifts the earlier result toward our prediction.

In closing we reiterate that the pion's valence-quark parton distribution amplitude does not express all bound-state dynamics associated with the pion.  Definitive features of Goldstone boson structure are also displayed in the pseudoscalar projection of the pion's Poincar\'e-covariant Bethe-Salpeter amplitude onto the light-front, which, according to a modern hypothesis, images the light-front distribution of the chiral condensate.  In this connection, our analysis provides a model-independent demonstration that this  condensate is primarily located in components of the pion's wave-function that express correlations with large relative momenta, a feature which ensures, \emph{inter alia},  that light-front longitudinal zero modes do not play a material role in forming this condensate.
\smallskip

\noindent\textbf{Acknowledgments}.
We are grateful for insightful comments from S.\,J.~Brodsky, I.\,C.~Clo\"et and P.\,C.~Tandy.
Work supported by:
For\-schungs\-zentrum J\"ulich GmbH;
and Department of Energy, Office of Nuclear Physics, contract no.~DE-AC02-06CH11357.
%



\begin{thebibliography}{59}
\expandafter\ifx\csname natexlab\endcsname\relax\def\natexlab#1{#1}\fi
\providecommand{\bibinfo}[2]{#2}
\ifx\xfnm\relax \def\xfnm[#1]{\unskip,\space#1}\fi
\bibitem[{Ward(1950)}]{Ward:1950xp}
\bibinfo{author}{J.~C. Ward}, \bibinfo{journal}{Phys. Rev.}
  \bibinfo{volume}{78} (\bibinfo{year}{1950}) \bibinfo{pages}{182}.
\bibitem[{Green(1953)}]{Green:1953te}
\bibinfo{author}{H.~S. Green}, \bibinfo{journal}{Proc. Phys. Soc. A}
  \bibinfo{volume}{66} (\bibinfo{year}{1953}) \bibinfo{pages}{873--880}.
\bibitem[{Takahashi(1957)}]{Takahashi:1957xn}
\bibinfo{author}{Y.~Takahashi}, \bibinfo{journal}{Nuovo Cim.}
  \bibinfo{volume}{6} (\bibinfo{year}{1957}) \bibinfo{pages}{371}.
\bibitem[{Collins et~al.(1977)Collins, Duncan, and Joglekar}]{Collins:1976yq}
\bibinfo{author}{J.~C. Collins}, \bibinfo{author}{A.~Duncan},
  \bibinfo{author}{S.~D. Joglekar}, \bibinfo{journal}{Phys. Rev. D}
  \bibinfo{volume}{16} (\bibinfo{year}{1977}) \bibinfo{pages}{438--449}.
\bibitem[{Nielsen(1977)}]{Nielsen:1977sy}
\bibinfo{author}{N.~Nielsen}, \bibinfo{journal}{Nucl. Phys. B}
  \bibinfo{volume}{120} (\bibinfo{year}{1977}) \bibinfo{pages}{212--220}.
\bibitem[{Pascual and Tarrach(1984)}]{tarrach}
\bibinfo{author}{P.~Pascual}, \bibinfo{author}{R.~Tarrach},
  \bibinfo{title}{QCD: Renormalization for the Practitioner},
  \bibinfo{publisher}{Springer-Verlag, Berlin}, \bibinfo{year}{1984}.
  \bibinfo{note}{{L}ecture Notes in Physics \textbf{194}}.
\bibitem[{Brodsky et~al.(1980)Brodsky, Frishman, Lepage, and
  Sachrajda}]{Brodsky:1980ny}
\bibinfo{author}{S.~J. Brodsky}, \bibinfo{author}{Y.~Frishman},
  \bibinfo{author}{G.~P. Lepage}, \bibinfo{author}{C.~T. Sachrajda},
  \bibinfo{journal}{Phys. Lett. B} \bibinfo{volume}{91} (\bibinfo{year}{1980})
  \bibinfo{pages}{239}.
\bibitem[{Braun et~al.(2003)Braun, Korchemsky, and Mueller}]{Braun:2003rp}
\bibinfo{author}{V.~Braun}, \bibinfo{author}{G.~Korchemsky},
  \bibinfo{author}{D.~Mueller}, \bibinfo{journal}{Prog. Part. Nucl. Phys.}
  \bibinfo{volume}{51} (\bibinfo{year}{2003}) \bibinfo{pages}{311--398}.
\bibitem[{Choi and Ji(2008)}]{Choi:2008yj}
\bibinfo{author}{H.-M. Choi}, \bibinfo{author}{C.-R. Ji},
  \bibinfo{journal}{Phys. Rev. D} \bibinfo{volume}{77} (\bibinfo{year}{2008})
  \bibinfo{pages}{113004}.
\bibitem[{Grigoryan and Radyushkin(2008)}]{Grigoryan:2008cc}
\bibinfo{author}{H.~R. Grigoryan}, \bibinfo{author}{A.~V. Radyushkin},
  \bibinfo{journal}{Phys. Rev. D} \bibinfo{volume}{78} (\bibinfo{year}{2008})
  \bibinfo{pages}{115008}.
\bibitem[{Chabysheva and Hiller(p ph)}]{Chabysheva:2012fe}
\bibinfo{author}{S.~Chabysheva}, \bibinfo{author}{J.~Hiller}
  (\bibinfo{year}{arXiv:1207.7128 [hep-ph]}). \bibinfo{note}{{\emph{A Dynamical
  Model for Longitudinal Wave Functions in Light-Front Holographic QCD}}}.
\bibitem[{Brodsky et~al.(p ph)Brodsky, de~Téramond, and
  Dosch}]{Brodsky:2013npa}
\bibinfo{author}{S.~J. Brodsky}, \bibinfo{author}{G.~F. de~Téramond},
  \bibinfo{author}{H.~G. Dosch}  (\bibinfo{year}{arXiv:1302.5399 [hep-ph]}).
  \bibinfo{note}{{\emph{Conformal Symmetry, Confinement, and Light-Front
  Holographic QCD}}}.
\bibitem[{Adler et~al.(1977)Adler, Collins, and Duncan}]{Adler:1976zt}
\bibinfo{author}{S.~L. Adler}, \bibinfo{author}{J.~C. Collins},
  \bibinfo{author}{A.~Duncan}, \bibinfo{journal}{Phys. Rev. D}
  \bibinfo{volume}{15} (\bibinfo{year}{1977}) \bibinfo{pages}{1712}.
\bibitem[{Roberts(2008)}]{Roberts:2007ji}
\bibinfo{author}{C.~D. Roberts}, \bibinfo{journal}{Prog. Part. Nucl. Phys.}
  \bibinfo{volume}{61} (\bibinfo{year}{2008}) \bibinfo{pages}{50--65}.
\bibitem[{Braun and Janot(2011)}]{Braun:2011fw}
\bibinfo{author}{J.~Braun}, \bibinfo{author}{A.~Janot}, \bibinfo{journal}{Phys.
  Rev. D} \bibinfo{volume}{84} (\bibinfo{year}{2011}) \bibinfo{pages}{114022}.
\bibitem[{Bashir et~al.(2012)Bashir, Chang, Cloet, El-Bennich, Liu
  et~al.}]{Bashir:2012fs}
\bibinfo{author}{A.~Bashir}, \bibinfo{author}{L.~Chang}, \bibinfo{author}{I.~C.
  Cloet}, \bibinfo{author}{B.~El-Bennich}, \bibinfo{author}{Y.-x. Liu}, et~al.,
  \bibinfo{journal}{Commun. Theor. Phys.} \bibinfo{volume}{58}
  (\bibinfo{year}{2012}) \bibinfo{pages}{79--134}.
\bibitem[{Petreczky(2012)}]{Petreczky:2012rq}
\bibinfo{author}{P.~Petreczky}, \bibinfo{journal}{J. Phys. G}
  \bibinfo{volume}{39} (\bibinfo{year}{2012}) \bibinfo{pages}{093002}.
\bibitem[{Fischer and Luecker(2013)}]{Fischer:2012vc}
\bibinfo{author}{C.~S. Fischer}, \bibinfo{author}{J.~Luecker},
  \bibinfo{journal}{Phys. Lett. B} \bibinfo{volume}{718} (\bibinfo{year}{2013})
  \bibinfo{pages}{1036--1043}.
\bibitem[{Roberts(l th)}]{Roberts:2012sv}
\bibinfo{author}{C.~D. Roberts}  (\bibinfo{year}{arXiv:1203.5341 [nucl-th]}).
  \bibinfo{note}{{{\emph{Strong QCD and Dyson-Schwinger Equations}}}}.
\bibitem[{{The Committee on the Assessment of and Outlook for Nuclear Physics;
  Board on Physics and Astronomy; Division on Engineering and Physical
  Sciences; National Research Council}(2012)}]{national2012Nuclear}
\bibinfo{author}{{The Committee on the Assessment of and Outlook for Nuclear
  Physics; Board on Physics and Astronomy; Division on Engineering and Physical
  Sciences; National Research Council}}, \bibinfo{title}{Nuclear Physics:
  Exploring the Heart of Matter}, \bibinfo{publisher}{National Academies
  Press}, \bibinfo{year}{2012}.
\bibitem[{Chang and Roberts(2012)}]{Chang:2011ei}
\bibinfo{author}{L.~Chang}, \bibinfo{author}{C.~D. Roberts},
  \bibinfo{journal}{Phys. Rev. C} \bibinfo{volume}{85} (\bibinfo{year}{2012})
  \bibinfo{pages}{052201(R)}.
\bibitem[{Chen et~al.(2012)Chen, Chang, Roberts, Wan, and Wilson}]{Chen:2012qr}
\bibinfo{author}{C.~Chen}, \bibinfo{author}{L.~Chang}, \bibinfo{author}{C.~D.
  Roberts}, \bibinfo{author}{S.-L. Wan}, \bibinfo{author}{D.~J. Wilson},
  \bibinfo{journal}{Few Body Syst.} \bibinfo{volume}{53} (\bibinfo{year}{2012})
  \bibinfo{pages}{293--326}.
\bibitem[{Wilson et~al.(2012)Wilson, Clo{\"e}t, Chang, and
  Roberts}]{Wilson:2011aa}
\bibinfo{author}{D.~J. Wilson}, \bibinfo{author}{I.~C. Clo{\"e}t},
  \bibinfo{author}{L.~Chang}, \bibinfo{author}{C.~D. Roberts},
  \bibinfo{journal}{Phys. Rev. C} \bibinfo{volume}{85} (\bibinfo{year}{2012})
  \bibinfo{pages}{025205}.
\bibitem[{Clo{\"e}t et~al.(2013)Clo{\"e}t, Roberts, and Thomas}]{Cloet:2013gva}
\bibinfo{author}{I.~C. Clo{\"e}t}, \bibinfo{author}{C.~D. Roberts},
  \bibinfo{author}{A.~W. Thomas}, \bibinfo{journal}{Phys. Rev. Lett.}
  \bibinfo{volume}{111} (\bibinfo{year}{2013}) \bibinfo{pages}{101803}.
\bibitem[{Brodsky and Shrock(2008)}]{Brodsky:2008be}
\bibinfo{author}{S.~J. Brodsky}, \bibinfo{author}{R.~Shrock},
  \bibinfo{journal}{Phys. Lett. B} \bibinfo{volume}{666} (\bibinfo{year}{2008})
  \bibinfo{pages}{95--99}.
\bibitem[{Brodsky and Shrock(2011)}]{Brodsky:2009zd}
\bibinfo{author}{S.~J. Brodsky}, \bibinfo{author}{R.~Shrock},
  \bibinfo{journal}{Proc. Nat. Acad. Sci.} \bibinfo{volume}{108}
  (\bibinfo{year}{2011}) \bibinfo{pages}{45--50}.
\bibitem[{Brodsky et~al.(2010)Brodsky, Roberts, Shrock, and
  Tandy}]{Brodsky:2010xf}
\bibinfo{author}{S.~J. Brodsky}, \bibinfo{author}{C.~D. Roberts},
  \bibinfo{author}{R.~Shrock}, \bibinfo{author}{P.~C. Tandy},
  \bibinfo{journal}{Phys. Rev. C} \bibinfo{volume}{82} (\bibinfo{year}{2010})
  \bibinfo{pages}{022201(R)}.
\bibitem[{Glazek(2011)}]{Glazek:2011vg}
\bibinfo{author}{S.~D. Glazek}, \bibinfo{journal}{Acta Phys. Polon. B}
  \bibinfo{volume}{42} (\bibinfo{year}{2011}) \bibinfo{pages}{1933--2010}.
\bibitem[{Chang et~al.(2012)Chang, Roberts, and Tandy}]{Chang:2011mu}
\bibinfo{author}{L.~Chang}, \bibinfo{author}{C.~D. Roberts},
  \bibinfo{author}{P.~C. Tandy}, \bibinfo{journal}{Phys. Rev. C}
  \bibinfo{volume}{85} (\bibinfo{year}{2012}) \bibinfo{pages}{012201(R)}.
\bibitem[{Brodsky et~al.(2012)Brodsky, Roberts, Shrock, and
  Tandy}]{Brodsky:2012ku}
\bibinfo{author}{S.~J. Brodsky}, \bibinfo{author}{C.~D. Roberts},
  \bibinfo{author}{R.~Shrock}, \bibinfo{author}{P.~C. Tandy},
  \bibinfo{journal}{Phys. Rev. C} \bibinfo{volume}{85} (\bibinfo{year}{2012})
  \bibinfo{pages}{065202}.
\bibitem[{Maris et~al.(1998)Maris, Roberts, and Tandy}]{Maris:1997hd}
\bibinfo{author}{P.~Maris}, \bibinfo{author}{C.~D. Roberts},
  \bibinfo{author}{P.~C. Tandy}, \bibinfo{journal}{Phys. Lett. B}
  \bibinfo{volume}{420} (\bibinfo{year}{1998}) \bibinfo{pages}{267--273}.
\bibitem[{H{\"o}ll et~al.(2004)H{\"o}ll, Krassnigg, and Roberts}]{Holl:2004fr}
\bibinfo{author}{A.~H{\"o}ll}, \bibinfo{author}{A.~Krassnigg},
  \bibinfo{author}{C.~D. Roberts}, \bibinfo{journal}{Phys. Rev. C}
  \bibinfo{volume}{70} (\bibinfo{year}{2004}) \bibinfo{pages}{042203(R)}.
\bibitem[{Keister and Polyzou(1991)}]{Keister:1991sb}
\bibinfo{author}{B.~D. Keister}, \bibinfo{author}{W.~N. Polyzou},
  \bibinfo{journal}{Adv. Nucl. Phys.} \bibinfo{volume}{20}
  (\bibinfo{year}{1991}) \bibinfo{pages}{225--479}.
\bibitem[{Brodsky et~al.(1998)Brodsky, Pauli, and Pinsky}]{Brodsky:1997de}
\bibinfo{author}{S.~J. Brodsky}, \bibinfo{author}{H.-C. Pauli},
  \bibinfo{author}{S.~S. Pinsky}, \bibinfo{journal}{Phys. Rept.}
  \bibinfo{volume}{301} (\bibinfo{year}{1998}) \bibinfo{pages}{299--486}.
\bibitem[{Chang et~al.(2013)Chang, Clo{\"e}t, Cobos-Martinez, Roberts, Schmidt
  et~al.}]{Chang:2013pq}
\bibinfo{author}{L.~Chang}, \bibinfo{author}{I.~C. Clo{\"e}t},
  \bibinfo{author}{J.~J. Cobos-Martinez}, \bibinfo{author}{C.~D. Roberts},
  \bibinfo{author}{S.~M. Schmidt}, et~al., \bibinfo{journal}{Phys. Rev. Lett.}
  \bibinfo{volume}{110} (\bibinfo{year}{2013}) \bibinfo{pages}{132001}.
\bibitem[{Lepage and Brodsky(1980)}]{Lepage:1980fj}
\bibinfo{author}{G.~P. Lepage}, \bibinfo{author}{S.~J. Brodsky},
  \bibinfo{journal}{Phys. Rev. D} \bibinfo{volume}{22} (\bibinfo{year}{1980})
  \bibinfo{pages}{2157}.
\bibitem[{Casher and Susskind(1974)}]{Casher:1974xd}
\bibinfo{author}{A.~Casher}, \bibinfo{author}{L.~Susskind},
  \bibinfo{journal}{Phys. Rev.} \bibinfo{volume}{D9} (\bibinfo{year}{1974})
  \bibinfo{pages}{436--460}.
\bibitem[{Braun and Filyanov(1990)}]{Braun:1989iv}
\bibinfo{author}{V.~M. Braun}, \bibinfo{author}{I.~Filyanov},
  \bibinfo{journal}{Z. Phys. C} \bibinfo{volume}{48} (\bibinfo{year}{1990})
  \bibinfo{pages}{239--248}.
\bibitem[{Beneke and Neubert(2003)}]{Beneke:2003zv}
\bibinfo{author}{M.~Beneke}, \bibinfo{author}{M.~Neubert},
  \bibinfo{journal}{Nucl. Phys. B} \bibinfo{volume}{675} (\bibinfo{year}{2003})
  \bibinfo{pages}{333--415}.
\bibitem[{Ball(1999)}]{Ball:1998je}
\bibinfo{author}{P.~Ball}, \bibinfo{journal}{JHEP} \bibinfo{volume}{9901}
  (\bibinfo{year}{1999}) \bibinfo{pages}{010}.
\bibitem[{Ball et~al.(2006)Ball, Braun, and Lenz}]{Ball:2006wn}
\bibinfo{author}{P.~Ball}, \bibinfo{author}{V.~Braun},
  \bibinfo{author}{A.~Lenz}, \bibinfo{journal}{JHEP} \bibinfo{volume}{0605}
  (\bibinfo{year}{2006}) \bibinfo{pages}{004}.
\bibitem[{Bowman et~al.(2004)}]{Bowman:2004jm}
\bibinfo{author}{P.~O. Bowman}, et~al., \bibinfo{journal}{Phys. Rev. D}
  \bibinfo{volume}{70} (\bibinfo{year}{2004}) \bibinfo{pages}{034509}.
\bibitem[{Cucchieri et~al.(2012)Cucchieri, Dudal, Mendes, and
  Vandersickel}]{Cucchieri:2011ig}
\bibinfo{author}{A.~Cucchieri}, \bibinfo{author}{D.~Dudal},
  \bibinfo{author}{T.~Mendes}, \bibinfo{author}{N.~Vandersickel},
  \bibinfo{journal}{Phys. Rev. D} \bibinfo{volume}{85} (\bibinfo{year}{2012})
  \bibinfo{pages}{094513}.
\bibitem[{Boucaud et~al.(2012)Boucaud, Leroy, Yaouanc, Micheli, Pene
  et~al.}]{Boucaud:2011ug}
\bibinfo{author}{P.~Boucaud}, \bibinfo{author}{J.~Leroy},
  \bibinfo{author}{A.~L. Yaouanc}, \bibinfo{author}{J.~Micheli},
  \bibinfo{author}{O.~Pene}, et~al., \bibinfo{journal}{Few Body Syst.}
  \bibinfo{volume}{53} (\bibinfo{year}{2012}) \bibinfo{pages}{387--436}.
\bibitem[{Ayala et~al.(2012)Ayala, Bashir, Binosi, Cristoforetti, and
  Rodriguez-Quintero}]{Ayala:2012pb}
\bibinfo{author}{A.~Ayala}, \bibinfo{author}{A.~Bashir},
  \bibinfo{author}{D.~Binosi}, \bibinfo{author}{M.~Cristoforetti},
  \bibinfo{author}{J.~Rodriguez-Quintero}, \bibinfo{journal}{Phys. Rev. D}
  \bibinfo{volume}{86} (\bibinfo{year}{2012}) \bibinfo{pages}{074512}.
\bibitem[{Aguilar et~al.(2012)Aguilar, Binosi, and
  Papavassiliou}]{Aguilar:2012rz}
\bibinfo{author}{A.~Aguilar}, \bibinfo{author}{D.~Binosi},
  \bibinfo{author}{J.~Papavassiliou}, \bibinfo{journal}{Phys. Rev. D}
  \bibinfo{volume}{86} (\bibinfo{year}{2012}) \bibinfo{pages}{014032}.
\bibitem[{Strauss et~al.(2012)Strauss, Fischer, and
  Kellermann}]{Strauss:2012dg}
\bibinfo{author}{S.~Strauss}, \bibinfo{author}{C.~S. Fischer},
  \bibinfo{author}{C.~Kellermann}, \bibinfo{journal}{Phys. Rev. Lett.}
  \bibinfo{volume}{109} (\bibinfo{year}{2012}) \bibinfo{pages}{252001}.
\bibitem[{Clo{\"e}t et~al.(2013)Clo{\"e}t, Chang, Roberts, Schmidt, and
  Tandy}]{Cloet:2013tta}
\bibinfo{author}{I.~C. Clo{\"e}t}, \bibinfo{author}{L.~Chang},
  \bibinfo{author}{C.~D. Roberts}, \bibinfo{author}{S.~M. Schmidt},
  \bibinfo{author}{P.~C. Tandy}, \bibinfo{journal}{Phys. Rev. Lett.}
  \bibinfo{volume}{111} (\bibinfo{year}{2013}) \bibinfo{pages}{092001}.
\bibitem[{Maris and Roberts(1997)}]{Maris:1997tm}
\bibinfo{author}{P.~Maris}, \bibinfo{author}{C.~D. Roberts},
  \bibinfo{journal}{Phys. Rev. C} \bibinfo{volume}{56} (\bibinfo{year}{1997})
  \bibinfo{pages}{3369--3383}.
\bibitem[{Qin et~al.(2012)Qin, Chang, Liu, Roberts, and Wilson}]{Qin:2011xq}
\bibinfo{author}{S.-x. Qin}, \bibinfo{author}{L.~Chang}, \bibinfo{author}{Y.-x.
  Liu}, \bibinfo{author}{C.~D. Roberts}, \bibinfo{author}{D.~J. Wilson},
  \bibinfo{journal}{Phys. Rev. C} \bibinfo{volume}{85} (\bibinfo{year}{2012})
  \bibinfo{pages}{035202}.
\bibitem[{Lane(1974)}]{Lane:1974he}
\bibinfo{author}{K.~D. Lane}, \bibinfo{journal}{Phys. Rev. D}
  \bibinfo{volume}{10} (\bibinfo{year}{1974}) \bibinfo{pages}{2605}.
\bibitem[{Politzer(1976)}]{Politzer:1976tv}
\bibinfo{author}{H.~D. Politzer}, \bibinfo{journal}{Nucl. Phys.}
  \bibinfo{volume}{B117} (\bibinfo{year}{1976}) \bibinfo{pages}{397}.
\bibitem[{Langfeld et~al.(2003)Langfeld, Markum, Pullirsch, Roberts, and
  Schmidt}]{Langfeld:2003ye}
\bibinfo{author}{K.~Langfeld}, \bibinfo{author}{H.~Markum},
  \bibinfo{author}{R.~Pullirsch}, \bibinfo{author}{C.~D. Roberts},
  \bibinfo{author}{S.~M. Schmidt}, \bibinfo{journal}{Phys. Rev. C}
  \bibinfo{volume}{67} (\bibinfo{year}{2003}) \bibinfo{pages}{065206}.
\bibitem[{Qin et~al.(2011)Qin, Chang, Liu, Roberts, and Wilson}]{Qin:2011dd}
\bibinfo{author}{S.-x. Qin}, \bibinfo{author}{L.~Chang}, \bibinfo{author}{Y.-x.
  Liu}, \bibinfo{author}{C.~D. Roberts}, \bibinfo{author}{D.~J. Wilson},
  \bibinfo{journal}{Phys. Rev. C} \bibinfo{volume}{84} (\bibinfo{year}{2011})
  \bibinfo{pages}{042202(R)}.
\bibitem[{Skullerud et~al.(2003)Skullerud, Bowman, K{\i}z{\i}lers{\"u},
  Leinweber, and Williams}]{Skullerud:2003qu}
\bibinfo{author}{J.~I. Skullerud}, \bibinfo{author}{P.~O. Bowman},
  \bibinfo{author}{A.~K{\i}z{\i}lers{\"u}}, \bibinfo{author}{D.~B. Leinweber},
  \bibinfo{author}{A.~G. Williams}, \bibinfo{journal}{JHEP}
  \bibinfo{volume}{04} (\bibinfo{year}{2003}) \bibinfo{pages}{047}.
\bibitem[{Bhagwat and Tandy(2004)}]{Bhagwat:2004kj}
\bibinfo{author}{M.~S. Bhagwat}, \bibinfo{author}{P.~C. Tandy},
  \bibinfo{journal}{Phys. Rev. D} \bibinfo{volume}{70} (\bibinfo{year}{2004})
  \bibinfo{pages}{094039}.
\bibitem[{Chang et~al.(2013)Chang, Roberts, and Schmidt}]{Chang:2012cc}
\bibinfo{author}{L.~Chang}, \bibinfo{author}{C.~D. Roberts},
  \bibinfo{author}{S.~M. Schmidt}, \bibinfo{journal}{Phys. Rev. C}
  \bibinfo{volume}{87} (\bibinfo{year}{2013}) \bibinfo{pages}{015203}.
\bibitem[{Chang and Roberts(2009)}]{Chang:2009zb}
\bibinfo{author}{L.~Chang}, \bibinfo{author}{C.~D. Roberts},
  \bibinfo{journal}{Phys. Rev. Lett.} \bibinfo{volume}{103}
  (\bibinfo{year}{2009}) \bibinfo{pages}{081601}.
\bibitem[{Nakanishi(1963)}]{Nakanishi:1963zz}
\bibinfo{author}{N.~Nakanishi}, \bibinfo{journal}{Phys. Rev.}
  \bibinfo{volume}{130} (\bibinfo{year}{1963}) \bibinfo{pages}{1230--1235}.

\end{thebibliography}

\end{document}